\documentclass[twocolumn,showpacs,showkeys,preprintnumbers,amsmath,amssymb]{revtex4}

\usepackage{epsfig}
\usepackage{graphicx}
\usepackage{amsmath}
\usepackage{graphicx}
\usepackage{dcolumn}
\usepackage{bm}
\usepackage{subfigure}%

\begin{document} 

\title[Mesoscopic model of waterlike fluids with 
hydrodynamic interactions]
{Towards a mesoscopic model of water-like fluids with hydrodynamic interactions}
\author{Irene Mazzitelli} \email{irene.mazzitelli@gmail.com}
\affiliation{IAC-CNR, via dei Taurini 9, 00185, Roma, Italy}
\author{Maddalena Venturoli} \email{maddalena.venturoli@roma1.infn.it}
\affiliation{Physics Department, University of Rome ``La Sapienza'', Piazzale A. Moro, 5, 00185, Roma, Italy}
\author{Simone Melchionna} \email{simone.melchionna@roma1.infn.it}
\affiliation{IPCF-CNR, Piazzale A. Moro, 5, 00185, Roma, Italy}
\author{Sauro Succi} \email{succi@iac.cnr.it}
\affiliation{IAC-CNR, via dei Taurini 9, 00185, Roma, Italy and Freiburg Institute
  for Advanced Studies, University of Freiburg,  Albertstrasse 19, 79104, Freiburg, Germany}

\date{\today}

\begin{abstract}
  We present a mesoscopic lattice model for non-ideal fluid flows with
  directional interactions, mimicking the effects of hydrogen-bonds in
  water.  The model supports a rich and complex structural dynamics of
  the orientational order parameter, and exhibits the formation of
  disordered domains whose size and shape depend on the relative
  strength of directional order and thermal diffusivity. By letting the
  directional forces carry an inverse density dependence, the model is
  able to display a correlation between ordered domains and low density
  regions, reflecting the idea of water as a denser liquid in the
  disordered state than in the ordered one.
\end{abstract}


\maketitle

\clearpage
\newpage

\section{Introduction}
\label{sec:intro}
Water is a most common fluid, and yet one still full of mysteries.
Indeed, water has been reckoned to exhibit dozens of anomalies, as
compared to standard fluids, primarily the fact of being denser in the
liquid than solid phase, exhibiting a density maximum at $4^oC$,
i.e. above the freezing point (for a vivid non technical description,
see \cite{Cart10}).  Although a fully comprehensive theory of water
thermodynamics is still missing, there is an increasing consensus that
most of these anomalies can be traced back to the peculiar nature of
the hydrogen bond (HB).  The HB interaction plays a vital role on
structure formation within water.  For instance, in water at low
temperature, the HB's lead to the formation of an open, approximately
four-coordinated (tetrahedral) structure, in which entropy, internal
energy and density decrease with decreasing temperature
\cite{Poole94}.  The equilibrium thermodynamics, i.e. phase diagram,
of water is exceedingly rich, and an ab-initio comprehensive analysis
of its properties is beyond computational reach.  As a result, many
models have been developed \cite{Dill}, including lattice ones, which
display {\it water-like} behavior \cite{LGW,Sastry93,Franzese10}.
Such lattice models are typically based on a many-body lattice-gas
Hamiltonian mimicking the essential features of water interactions,
with no claim/aim of/at atomistic fidelity \cite{LGW2}.  To the best
of our knowledge, these models have been employed mostly for the study
of equilibrium properties, typically via Monte Carlo simulations.
Yet, in most phenomena of practical interest, water flows and, most
importantly, a variety of molecules, say colloids, ions and
biopolymers, flow along with it, typically in nanoconfined geometries.
In the biological context, it is well known that the competition
between hydrophobic and hydrophilic interactions plays a crucial role
in affecting the conformational dynamics of proteins \cite{Karplus,
Ciepa, Rao}.  On a larger scale, hydrodynamic interactions are know to
exert a significant effect on the collective dynamics and aggregation
phenomena within protein suspensions. More generally, hydrodynamic
interactions are crucial in the presence of confining walls, due to
their strong coupling with resulting inhomogeneities \cite{PhysTod}.

Based on the above, there is clearly wide scope for a minimal model of
water behavior, capable of including hydrodynamic interactions and
geometrical confinement, at a mesoscopic level (say tens of nanometers
to tens of microns).  In this respect, a remarkable mesoscopic
methodology has emerged in the last two decades, in the form of
minimal versions of the (lattice) Boltzmann kinetic equation
\cite{ben92,che98,wg00,succi01}.  Such lattice Boltzmann equations
(LBE's) have proven fairly successful in simulating a broad variety of
complex flows across scales, from macroscopic fluid turbulence, all
the way down to biopolymer translocation in nanopores \cite{STATPHYS}.
The LB approach is mostly valued for its flexibility towards the
treatment of complex geometries and seamless inclusion of complex
physical interactions, e.g. flows with phase transitions.  Such
advantages are only accrued by the outstanding computational
efficiency of the method, especially on parallel computers
\cite{AMATI, COVEN}.  To the best of our knowledge, however, no LB
model for water-like fluids has been developed as yet.  In this paper,
we present the first preliminary effort to fill this gap.  More
specifically, we develop a new LB framework including HB-like
interactions, and explore the collective dynamics of the mesoscopic
water-like fluids, both in free space (homogeneous) and nano-confined
(heterogeneous) environments.

This paper is organized as follows.  In Sec.~\ref{sec:model}  
we present the basic elements of our mesoscopic approach, namely: 
in Sec.~\ref{sec:advection} we state the 
transport equations as described by the Lattice Boltzmann model for non-ideal
fluids, in Sec.~\ref{sec:DI} we describe the directional interactions mimicking hydrogen-bonds 
and in Sec.~\ref{sec:orderpar} we present the  model for the dynamics of the orientational order parameter.  
The details of the numerical scheme are provided in Sec~\ref{sec:scheme}.
Our results are presented in Sec.~\ref{sec:results} where
we first give a qualitative analysis of the mesoscopic
model (Sec.~\ref{sec:analysis}) and an estimate of the numerical parameters to be used in the
simulations (Sec.~\ref{sec:params}). We then present numerical results 
for homogeneous (Sec.~\ref{sec:homogeneous}) and inhomogeneous
(wall-confined) scenarios (Sec.~\ref{sec:heterogeneous} and \ref{sec:flow}), 
without and with hydrodynamic flow.
Finally, in Sec.~\ref{sec:outlook} we provide an outlook of possible directions of
future investigation.

\section{Mesoscopic model for water}
\label{sec:model}
\subsection{Transport equations}
\label{sec:advection}
\begin{figure}[htbp]
  \includegraphics[width=6.5cm]{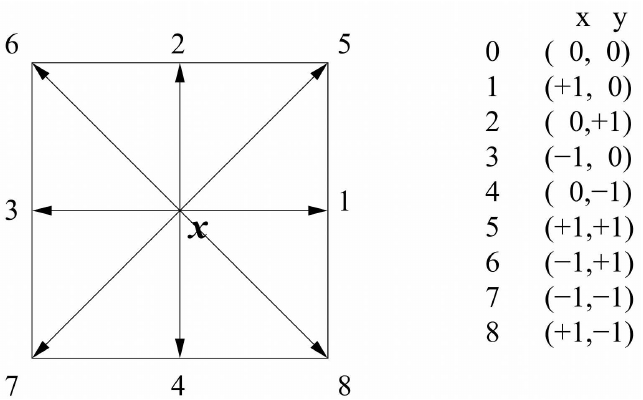}
  \caption{Distribution of the discrete molecular velocities 
    ${\bf c}_i$, $i=0,..,8$, in the two-dimensional D2Q9 lattice.}
  \label{fig:fig1}
 \end{figure}

Our fluid model is based on an extension 
of the lattice Boltzmann method for ideal fluids \cite{wg00,succi01}.
We use a two dimensional (2d) model on the D2Q9 lattice depicted in 
Fig.~\ref{fig:fig1}.
At each grid node ${\bf x}$
the velocity distribution function $f_i({\bf x},t)$, i.e. 
the probability to find a particle at location ${\bf x}$, 
moving along the lattice direction defined by the discrete 
speed ${\bf c}_i$, is evolved according to 
the kinetic equation, with Bhatnagar-Gross-Krook (BGK)
approximation \cite{bgk54,qia92}:
\begin{equation}
  f_i({\bf x} + {\bf c}_i,t+\Delta t)-f_i({\bf x},t)=
  -\omega \,[ f_i({\bf x},t)-f_i^{eq}({\bf x},t) ]
\label{kinetic}
\end{equation}
where $\omega = \Delta t/\tau$, with the time step $\Delta t =1$, and
$\tau$ the relaxation time towards local equilibrium.  The relaxation
time $\tau$ fixes the fluid kinematic viscosity $\nu = c_s^2(\tau
-1/2)$, where $c_s$ is the sound speed of the lattice fluid. Here we
have taken the mesh spacing $\Delta x=1$ so that $c_s = 1/\sqrt{3}$.

The local equilibrium distribution function is a Maxwellian expanded
to the second order in the fluid velocity \cite{wg00,succi01} and it
is described by the distribution functions $f_i^{eq}$
\begin{equation}
  \label{eq:feq}
  f_i^{eq} = 
  \begin{cases}
    w_i \rho \Big[ 1 - \frac{3{\bf U}^2}{2\,c^2} \Big] 
    \quad i=0 \\
    w_i \rho \Big[ 1 + 3 \frac{{\bf c}_i \cdot {\bf U}}{c^2} +
    \frac{9({\bf c}_i \cdot {\bf U})^2}{2\,c^4}
    - \frac{3{\bf U}^2}{2\,c^2} \Big] 
    \quad i=1,\ldots,8
  \end{cases}
\end{equation}
The weights $w_i$ for the D2Q9 lattice are 
\begin{equation}
  \label{eq:weights}
  w_i = 
  \begin{cases}
    \frac{4}{9}  ~\quad i=0 \\
    \frac{1}{9}  ~\quad i=1,2,3,4 \\
    \frac{1}{36} \quad i=5,6,7,8 \\
  \end{cases}
\end{equation}
and we take the propagation speed on the lattice $c=1$.  The
macroscopic variables in Eq.~\eqref{eq:feq} are the fluid density
$\rho$, and the fluid velocity ${\bf U}$, defined as follows:
\begin{equation}
  \rho = \rho({\bf x},t) = \sum_{i=0}^8 f_i({\bf x},t)
  \label{density}
\end{equation}
and:
\begin{equation}
  {\bf U} = {\bf U}({\bf x},t) = {\bf u}({\bf x},t) + {\tau \over 
    \rho({\bf x},t)}{\bf F}({\bf x},t) 
  \label{velocity}
\end{equation}
with 
\begin{equation}
  \rho({\bf x},t) {\bf u }({\bf x},t) = 
  \sum_{i=0}^8 f_i({\bf x},t){\bf c}_i.
  \label{momentum}
\end{equation}
The forcing term ${\bf F}({\bf x},t)$ in Eq.~(\ref{velocity}) reflects
the interparticle interactions.  At a microscopic (molecular) level,
these interactions are given by a combination of van der Waals and
electrostatic forces, which take into account excluded volume,
dispersion, directional hydrogen bonds and multipolar interactions.
In this work, the cohesive forces prevailing in the aqueous
environment are represented by a Shan-Chen pseudo-potential model
\cite{sha93}.  The water-water cohesive forces are taken proportional
to a free parameter ${\it G_b}$, and enter the momentum equations
(\ref{velocity}) via the forcing term
\begin{equation}
  {\bf F}({\bf x},t) = - G_b \psi({\bf x},t) \sum_{i=1}^{8} w_i 
  \psi({\bf x} + {\bf c}_{i},t) {\bf c}_{i}
  \label{eq:Fww}
\end{equation}
where the normalization weights $w_i$ are taken as in Eq.~\eqref{eq:weights}
and $\psi$ is a function
of the density, $\psi(\rho) = (1- \exp({-\rho})  ).$
Under these conditions, the fluid pressure 
receives a non-ideal contribution from potential energy interactions and
takes the form:
\begin{equation}
 p(\rho) = c_s^2\Big(\rho + {{\it G_b} \over 2} 
  \psi^2(\rho) \Big).
  \label{eq:press}
\end{equation}
This non-ideal equation of state supports a liquid-vapor  
phase transition for $G_b < -4$ (negative $G_b$ values code for attraction) at
a critical density $\rho_c = \ln 2$ in lattice units.
Note that hard-core, short-range repulsive interactions in charge of stabilizing
critical phase-separation are replaced by a self-driven saturation mechanism, whereby 
the force becomes vanishingly small as $\rho \gg 1$, as encoded in the exponential
dependence of the pseudo-potential $\psi$ on the density $\rho$,
and for a uniform density.

\subsection{Directional interactions}
\label{sec:DI}
The main distinctive feature of the present model consists in the inclusion of
{\it directional} interactions, aimed at mimicking hydrogen bonds at
a mesoscopic level.
At the molecular level, the main effect of hydrogen-bonding (HB) between 
different water molecules is to align a donor (hydrogen atom) of a 
given water molecule to the acceptor (oxygen atom)
of a neighboring molecule. The topology of these links is credited for exerting
very profound effects on the collective behavior of water, and ultimately
provides a basis for understanding its numerous anomalies.
In particular, in three dimensions, at low temperature HB's tend to favor 
tetrahedral structures (icy-water) which, being less packed than
isotropic molecules, would explain why ice (the ordered phase)
is less dense than liquid water (disordered phase).
HB's lead to the formation of complex and dynamic
network structures, fueled by the unceasing breaking/formation of new HB's.
Since these are very short-lived events, we cannot expect them 
to be detectable at a mesoscopic level.
At such a level, we expect to observe the dynamics
of a suitable order parameter, which takes a zero value 
in the disordered phase and non-zero in the ordered one, the latter
being promoted and sustained by mesoscopic directional interactions.
Likewise, at a molecular level, temperature takes the form of random noise, promoting
jumps between the different hydrogen-bond configurations (noise breaks the bond).
At a mesoscopic level, though, noise takes the connotation of a 
diffusive process, driving the system towards a mesoscopically uniform state.

 \begin{figure}[htbp]
   \includegraphics[width=3.2cm]{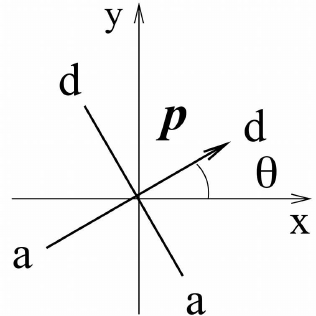}
   \caption{Model of a four-arm molecule.
     Two arms represents donors (d) and two arms are acceptors (a).
     The vector ${\bf p}=(p_x,p_y)$ is
     oriented in the direction of one donor, as shown in the figure.}
   \label{fig:fig2}
 \end{figure}

\begin{figure}[htbp]
   \includegraphics[width=6cm]{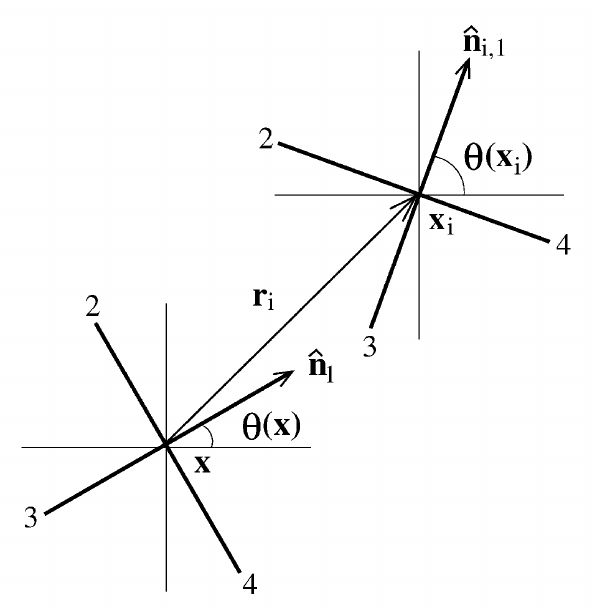}
   \caption{Sketch of two model ``molecules''
     interacting via hydrogen-bond potential.}
   \label{fig:fig3}
 \end{figure}

To incorporate directional interactions (DI), we endow
the lattice fluid with internal degrees
of freedom, in charge of responding to orientational forces.
These degrees of freedom are modeled in terms of four square-planar
oriented bonding arms, as illustrated in Fig.~\ref{fig:fig2}. 
The bonding arms of each ``molecule'' are either 
`donor'-like or `acceptor'-like, to indicate hydrogen or oxygen 
type behavior, respectively. 
Since the four arms are rigid, they are uniquely identified by a single
parameter, e.g.~the angle $\theta$ formed by, say, the 
first arm with the $x$ coordinate axis. 
However, such an angle is not a convenient order parameter
because of its periodicity.
As a result, we have introduced a vector order parameter 
${\bf p}=(p_x,p_y)$, pointing in the direction 
of one donor bonding arm (see Fig.~\ref{fig:fig2}).
The angle $\theta$ relates to the vector ${\bf p}$ 
through $\theta=\tan^{-1} (p_y/p_x)$.
The interaction between two molecules located, respectively, at
${\bf x}$ and ${\bf x}_i$, with ${\bf r}_i = {\bf x}_i -
{\bf x}$, is described by the following interparticle 
pseudo-potential \cite{dak08}:
\begin{equation}
  \begin{aligned}
    V_{HB}({\bf x},{\bf x}_{i}) &= \phi(\rho({\bf x})) \phi(\rho({\bf x}_i))  
     \exp \Big[ - {(r_{i} -R_{HB})^2 \over 2 
      \sigma^2_R}\Big]     \\ 
    \times \Big(  \sum_{k=1}^4 \sum_{k'=1}^4 &\epsilon_{HB}(k,k')  
    \exp \Big[ - \Big( { {\bf \hat{n}}_k\cdot {\bf r}_i \over r_i}-1    \Big)^2 
    {1 \over 2 \sigma^2_\theta} \Big]  \\
    &\times  \exp \Big[ - \Big( { {\bf \hat{n}}_{i,k'}\cdot {\bf r}_i \over r_i}  + 1 \Big)^2 
    { 1 \over 2 \sigma^2_\theta} \Big] \Big) 
  \end{aligned}
  \label{eq:VHB} 
\end{equation}
where $r_i=|{\bf r}_i|$ and ${\bf \hat{n}}_k$ (${\bf \hat{n}}_{i,k'}$), with $k\,(k')=1,..,4$, are 
the unit vectors  indicating the orientation of the four bonding arms of a molecule 
at grid node ${\bf x}$ (${\bf x}_i$) (see Fig.~\ref{fig:fig3}). 
$R_{HB}$ is the equilibrium radial distance of the HB's,
and $\sigma_R$ and $\sigma_\theta$ control the radial and angular
decay of the interactions around such equilibrium (stiffness).

The matrix $\epsilon_{HB}(k,k')$ is introduced to distinguish between 
donors (arms $1$ and $2$) and acceptors (arms $3$ and $4$). 
The interaction energy between donors and acceptors
is equal to the constant $\varepsilon_{HB}<0$, whereas the
interaction between two donors or two acceptors is set to zero. 

To include the density-dependent propensity of water to form
ordered states, which reflects the fact that tetrahedral ordered
structures are less compact than isotropic disordered ones (the two
liquid phases of water), we have introduced in Eq.~\eqref{eq:VHB}
the weight function $\phi(\rho)$   
\begin{equation}
 \phi(\rho)  = {1 \over 1+
    e^{-\alpha (\frac{\rho_{\rm max} - \rho}{\rho_{\rm max}-\rho_{\rm
          min}}) } }
  \label{eq:phiHB}
\end{equation}
where $\rho_{\rm max}$ and $\rho_{\rm min}$ are the maximum and
minimum density of the fluid under the chosen conditions and
$\alpha>0$ is a parameter controlling the range of the density
variation, which we have set equal to $10$ to implement a steeply
decaying function.

We can now further justify our choice of the vector order parameter,
${\bf p}$, by noting that the potential in Eq.~(\ref{eq:VHB}), with
$R_{HB}=\sqrt{2}$, has four minima, corresponding to
$\theta=\pi/4,3\pi/4,5\pi/4$ and $7\pi/4$.  As a result, $\cos\theta$
alone cannot distinguish between these four, leaving two of them
degenerate, unless $\sin\theta$ is also specified.  Clearly, the two
component vector ${\bf p}$ does not suffer of such limitation.

The magnitude of the total torque on 
${\bf p}({\bf x},t)$ due to the HB interaction with its eight
neighbors reads as follows:
\begin{equation}
  \Delta \theta ({\bf x},t) = 
  C_T \; \sum_{i=1}^8 c_i \Delta t F_i({\bf x},t)
  \label{totalT}
\end{equation}
where $C_T$ is a constant to be discussed shortly, $c_i =|{\bf c}_i|$, 
$\Delta t = 1$ and
$F_i({\bf x},t)$ is the angular component of the force between site
${\bf x}$ and ${\bf x}_i$, given by
\begin{equation}
  F_i({\bf x},t) = -\sum_{k=1}^4 { \partial V_{HB}({\bf x},{\bf x}_i) 
    \over c_i\partial \theta_k}
\end{equation}
where $\theta_k = \tan^{-1}(n_{ky}/n_{kx})$, $k=1,..,4$.
Thus
\begin{equation}
  F_i({\bf x},t) = -\sum_{k=1}^4  {\partial V_{HB}({\bf x},{\bf x}_i)  \over
    c_i \partial cos\alpha_k} { d cos\alpha_k \over d \theta_k}
  \label{fHB}
\end{equation}
with $\alpha_k$ the 
angle in between the unit vector in direction ${\bf \hat{n}}_k$
and the velocity ${\bf c}_i$
\begin{equation}
  cos\alpha_k = {{\bf \hat{n}}_k\cdot {\bf c}_{i} \over c_i} = {n_{kx} c_{ix} + n_{ky} c_{iy} \over  c_i};
  \label{cosalpha}
\end{equation}
here, $n_{kx} = cos\theta_k, \,\,\, n_{ky} = sin\theta_k$.
This gives
\begin{equation}
  \begin{aligned}
    \label{rotation}
    &c_i F_i({\bf x},t) = \phi(\rho({\bf x})) \phi(\rho({\bf x}_i)) 
    \exp \Big[ - {(c_{i} -R_{HB})^2 \over 2 \sigma^2_R} \Big]  \times \\
    &\Big( \sum_{k=1}^4  \sum_{k'=1}^4  \epsilon_{HB}(k,k') 
      2(\cos\alpha_k - 1) 
      \exp \Big[-\frac{(cos \alpha_k-1)^2}{2\sigma^2_{\theta}}\Big] \\
      &\times \Big[{-c_{ix} \sin\theta_k+c_{iy} \cos\theta_k \over c_i} }\Big] 
    {\exp \Big[ - \Big( {{\bf \hat{n}}_{i,k'}\cdot {\bf c}_i \over c_i}+1 \Big)^2 
      { 1 \over 2 \sigma^2_\theta} \Big] \Big). 
    \end{aligned}
  \end{equation}  

\subsection{Dynamics of the order parameter}
\label{sec:orderpar}
The idea underlying the present approach is that the mesoscopic
description of directional interactions is reflected by 
the hydrodynamic equation of the vector ${\bf p}$.
Such an equation must include three main effects: macroscopic advection,
bond formation due to directional interactions and bond-breaking due to thermal noise. 
The resulting transport equation reads as follows
\begin{equation}
  {\partial {\bf p} \over \partial t}  + {\bf u}\cdot \nabla {\bf p} = 
  {\bf T}  + {\bf D}.
  \label{equation_p}
\end{equation}
The two terms on the right hand side represent
the deterministic torque, ${\bf T}$, due to 
the hydrogen-bond interaction, and thermal diffusion, ${\bf D}$, 
due to translational motion.
As a result, the water-like fluid is characterized by 
the density $\rho$, the velocity ${\bf U}$ and the 
rotational vector ${\bf p}$.
The equation (\ref{equation_p}) is evolved concurrently with 
the LB equation (\ref{kinetic}) for the fluid density and velocity.

\subsubsection{Deterministic torque}
\label{sec:torque}
\begin{figure}[htbp]
   \includegraphics[width=3.5cm]{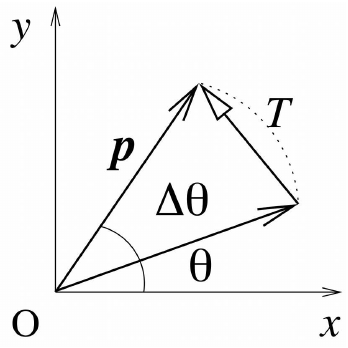}
   \caption{Deterministic rotation, ${\bf T}$, of the molecule whose
     initial orientation is $\theta 
     = tan^{-1} (p_y/p_x)$. Note that 
     $|{\bf T}| = 2|{\bf p}|\sin(|\Delta \theta|/2)$.}
   \label{fig:fig4}
 \end{figure}

The torque is represented by the force in Eq.~\eqref{rotation} which 
is inserted in the equations of motion for ${\bf p}$,
after projection of its components along the $x$ and $y$ directions. 
The two components can be computed as illustrated in Fig.~\ref{fig:fig4}. 
The molecule rotates from 
the initial orientation $\theta = tan^{-1}(p_y/p_x)$
to $\theta + \Delta \theta$. The angle $\Delta \theta$ is larger than $0$ for 
counterclockwise rotation, whereas $\Delta \theta <0$ for
clockwise rotation. 
By making use of suitable trigonometric relationships, one obtains:
\begin{eqnarray}
  \label{rotx}
  T_x &= - 2|{\bf p}|sin{\Delta \theta \over 2} sin\Big({\Delta \theta \over 2}
  +\theta \Big)\\
  \label{roty}
  T_y &= \  2|{\bf p}|sin{\Delta \theta \over 2} cos\Big({\Delta \theta \over 2}
  +\theta \Big).
\end{eqnarray}
As discussed previously, the torque is zero when the angle
$\theta$ is in one of the four degenerate minima $\theta_k = (2k-1) \pi/4$, $k=1,4$.

\subsubsection{Thermal diffusivity}
\label{sec:tdiff}
At a molecular scale, temperature acts as a HB-breaking noise which
promotes transitions between different bond configurations.  At a
mesoscopic level, such bond-breaking effect manifests itself as a
diffusion process, driving the system towards an isotropic, disordered
state $\langle \cos \theta \rangle = \langle \sin \theta \rangle = 0$,
where brackets stand for ensemble averaging over a mesoscopic volume
of fluid.  As a result, in our model thermal diffusion is represented
in standard laplacian form $ {\bf D} = D_p \Delta {\bf p}$, where
$D_p$ is the kinematic diffusivity of the vector ${\bf p}$, to be
detailed shortly.

\subsubsection{Hydrodynamic interactions}
\label{sec:hydro}
Hydrodynamic interactions are automatically included by moving the
vector ${\bf p}$ along with the fluid velocity ${\bf U}$, resulting
from the LB advection equation~\eqref{kinetic}.  These interactions
are expected to play a major role in transport phenomena, with the
fluid in motion and/or suspended bodies moving along with it.  In the
present study, however, we devote special attention to a simpler
scenario, namely the competition between directional interactions and
thermal diffusion.  Such competition is also analyzed in the presence
of hydrodynamic flows (Poiseuille), but mostly for illustrative
purposes. Quantitative investigation of the highly complex phenomena
resulting from the concurrent effects of directional interactions,
diffusion and hydrodynamic transport, are left to future studies.

\section{Numerical scheme for the order parameter}
\label{sec:scheme}
The equation~\eqref{equation_p} for ${\bf p}$ is essentially an
advection-diffusion-reaction equation, for which a wide variety of
numerical methods is available, in particular, the so-called, moment
propagation method, \cite{PAGO,PAGO2}.  For the sake of uniformity
with the advection equation described in Sec.~\ref{sec:advection}, and
also to secure very low numerical diffusivity, we have opted for a LB
integrator also for the equation of the order parameter.
Consequently, we integrate the equation of motion for ${\bf
p}=(p_x,p_y)$ by a lattice Boltzmann scheme, applied separately to the
two components of the vector.  This means that, using the same D2Q9
lattice geometry of Sec.~\ref{sec:advection}, we define two sets of
density distribution functions $g_{x,i}({\bf x},t)$ and $g_{y,i}({\bf
x},t)$, $i=0,..,8$, respectively for $p_x$ and $p_y$, such that
\begin{eqnarray}
  p_x = p_x({\bf x},t) = \sum_{i=0}^8 g_{x,i}({\bf x},t)
  \\
  p_x({\bf x},t) {\bf u}({\bf x},t) = \sum_{i=0}^8 g_{x,i}({\bf x},t) {\bf c}_i.
\end{eqnarray}
The kinetic equation for $g_{x,i}$ is:
\begin{equation}
  \begin{aligned}
    g_{x,i}({\bf x} + {\bf c}_i,t+1)-g_{x,i}({\bf x},t) &= \\
    -\omega_p(g_{x,i}({\bf x},t)  -& g_{x,i}^{eq}({\bf x},t)) + 
    T_i^x 
  \end{aligned}
  \label{kinetic_p}
\end{equation}
where $T_i^x=w_i T_x$ is the $x$ component of the
deterministic rotation, $\omega_p = 1/\tau_p$, with $\tau_p$ 
the relaxation time towards local equilibrium, described by the
density function
\begin{equation}
  g_{x,i}^{eq}=p_xw_i\Big[1 + {{\bf c}_i\cdot {\bf U} \over c_s^2} \Big], \quad \forall i.
  \label{equilibria_p}
\end{equation}
Similar expressions hold for $p_y$ by replacing $x$ with $y$. 

The hydrodynamic limit of this  LB model for ${\bf p}$ yields
the following continuum macroscopic equations:
\begin{equation}
  {{\partial {\bf p}} \over \partial t} + {\bf u} \cdot \nabla {\bf p} = {\bf T} + D_p \Delta {\bf p},
  \label{macroscopic_p}
\end{equation}
with $D_p = c_s^2(\tau_p-1/2)$. 
An appealing aspect of LB is that one can tune the diffusivity to very small values
by choosing 
\begin{equation}
  \tau_p = {1 \over 2} + \epsilon,
  \label{diffusivity}
\end{equation}
with $\epsilon$ typically of the order $1/N$, $N$ being the number 
of lattice sites per linear dimension.

We note that diffusivity is the emergent manifestation of microscopic
noise, and acts in such a way as to smear out spatial gradients of the
vector ${\bf p}$.  One could still add a stochastic source to the rhs
of equation (\ref{equation_p}), in order to model random noise. This,
however, would not be consistent with the mesoscopic aim of the
present model.
 
\section{Numerical results}
\label{sec:results}
\subsection{Analysis of the model}
\label{sec:analysis}
Before discussing the details of the simulation results, a few general
comments on the expected qualitative scenario are in order.  The
vector {\bf p} moves with the fluid and diffuses at a rate fixed by
the diffusivity $D_p$.  At the same time, it rotates under the effect
of the torque associated with directional interactions (DI) with the
neighboring fluid sites.  In the absence of DI's, and with the fluid
at rest ${\bf u}=0$, the order parameter would tend to a uniform state
(disorder), as dictated by thermal diffusivity.  DI's, on the other
hand, tend to place the system on local minima of the interaction
potential, thereby giving rise to metastable ordered domains.  The
torque, which depends only on angular degrees of freedom, takes
$\theta(x,y;t)$ to the local minimum closest to the initial condition
$\theta(x,y;t=0)$.  As a consequence, after an initial transient
stage, the system settles down into its local minima, with no
transitions between them.  Indeed, transitions between different
minima can be detected only in the initial stage, in which diffusion
is still capable of affecting the evolution of $\theta$ towards
different minima, because the system is still sufficiently far from
equilibrium.
Once this transient is over, the system freezes into a metastable
crystal-like state.  In this way, the system attains a state of
coexistence between short-range uniformity and long-range disorder.
Within each domain, the system remains uniform around the
corresponding local minimum angle.  On the other hand, the spatial
distribution of the domains has a fairly disordered pattern, depending
on the initial conditions and the (inverse) strength of the diffusion.
A richer scenario, i.e. a longer transient with more inter-domain
disorder, could be enforced by increasing the number of HB arms, so as
to enhance the number of angular minima, hence their mutual
competition.  This would give rise to a more disordered system, but
would not change once it falls into the closest minimum, the system
stays forever.

\subsection{Numerical estimate of the simulation parameters}
\label{sec:params}
The quantitative details of the scenario previously described depend
on the specific values of the parameters governing the physics of the
system, primarily the relative strength of DI's versus diffusion,
namely $\varepsilon_{HB}/kT$.  In the lattice Boltzmann model
$kT=1/3$.  In physical units, at ambient temperature $T=300K$, $kT
\simeq 2,3 KJ/mol$.  For a water dimer, the potential energy of the
hydrogen bond is of the order of $60KJ/mol$, whereas the free energy
can be estimated as $\Delta F \sim kT/2$ \cite{rao10}.  In the present
model, the role of the free energy, i.e. the energy difference between
bounded and unbounded states, is played by the parameter
$\varepsilon_{HB}$ in Eq.~(\ref{eq:VHB}).  Therefore, with
$\varepsilon_{HB} = -0.1$, we fulfill the condition $\varepsilon_{HB}
\sim \Delta F$.

The factor $C_T$, that, according to Eq.~(\ref{totalT}), weights the
effect of the torque, is estimated via the momentum equation
\begin{equation}
  I{d^2 \theta \over dt^2} = {\cal T} - I \gamma {d \theta \over dt}
  \label{inertia}
\end{equation}
where $I \simeq M a^2$ is the moment of inertia, $\gamma$ the drag
coefficient and ${\cal T}$ the torque. Thus, upon replacing
infinitesimal with finite increments, at steady state, we obtain
$\Delta \theta \simeq {{\cal T} \over I \gamma}\Delta t$.  With $M=1$
and $a=1$, this yields $C_T \propto 1/\gamma$ (see
Eq.~\eqref{totalT}).

Next, we estimate the relative effect of the torque with respect to
diffusion.  First of all, by Taylor expansion, close to the minimum of
${\cal T}$, Eq.~(\ref{inertia}) gives ${d \theta \over dt}= {{\cal T'}
\over I\gamma} \Delta \theta$ where ${\cal T'}\equiv (d{\cal
T}/d\theta)|_{\theta_{eq}}$.  Therefore, the strength of the
deterministic rotation can be quantified by the relaxation time
$\tau_T$ ${1 \over \tau_T } \equiv {{\cal T'} \over I\gamma}$.  On the
other hand, the diffusion time scale in Eq.~(\ref{equation_p}) can be
computed according to ${1 \over \tau_D} \sim {D_p \over w^2}$, with
$w$ the width of the interface, typically of the order of $5$ lattice
units in LB simulations.  The ratio between diffusive and
deterministic forcing is then given by
\begin{equation}
  {\tau_T \over \tau_D}
  ={ I \gamma D_p \over {\cal T'} {w}^2} = 
  { I \gamma D_p \sigma_\theta^2 \over |\varepsilon_{HB}| {w}^2}
  \label{ratiotau}
\end{equation}
where the second equality comes by dimensional analysis from
Eq.~(\ref{eq:VHB}), which leads to ${\cal T'} \simeq
|\varepsilon_{HB}| /\sigma_\theta^2$.  In our numerical simulations,
this ratio is changed by tuning two parameters, the strength of DI's
via $\varepsilon_{HB}$ and the diffusivity, via the parameter
$\epsilon$. Finally, we set $\gamma=1$ for numerical convenience.

\subsection{Homogeneous scenario}
\label{sec:homogeneous}
\begin{figure}[thbp]
   \includegraphics[width=3.25in]{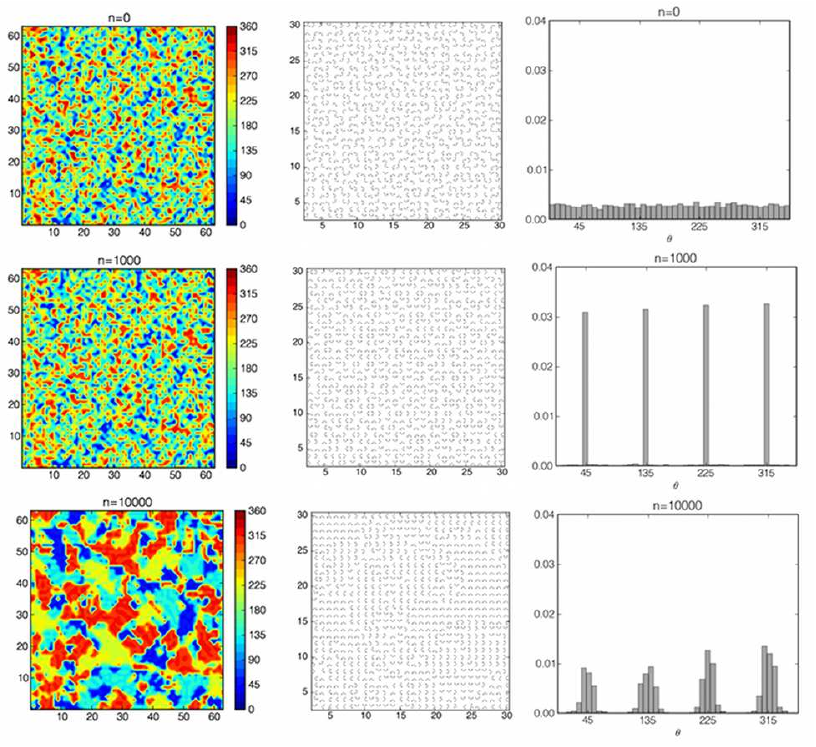}
   \caption{
     Configuration of $\theta$ in a $64^2$ homogeneous system, shown as
     contour plot (left), detail of the orientation of the four-armed
     ``molecules'' (middle, only donor arms are represented), and histogram
     (right).  Starting from the same initial ($n$=0) random configuration
     shown in the top panels, we report the equilibrium configurations in
     the following two cases; Central panels: Run (a), the scenario with
     $\epsilon_{HB}=-1\times10^{-1}$ and no diffusion in the dynamics of
     ${\bf p}$ (i.e. $\epsilon=0$).  Equilibrium is reached after $n
     \approx 10^3$ steps.  Bottom panels: Run (b), the scenario with
     $\epsilon_{HB}=-1\times10^{-1}$ and diffusion with
     $\epsilon=0.01$. Equilibrium is reached after $n\approx10^4$ steps.
     Note that in both cases the histograms are peaked around the four
     equilibrium values $\theta=\pi/4,3\pi/4,5\pi/4,7\pi/4$, but the size
     of the domains increases when diffusion is turned on.  Also, the peaks
     of the histogram broaden with increasing diffusion, indicating thicker
     domain boundaries.}
   \label{fig:fig5}
 \end{figure}

We first consider the case of a bulk fluid without density-dependent
interactions, i.e.~where $G_b=0$ and $\phi(\rho)=1$. The heterogeneous
case will be considered in Sec.~\ref{sec:heterogeneous}.  To study the
effect of the different contributions to the dynamics of the order
parameter, we start from the random initial condition shown in the top
panels of Fig.~\ref{fig:fig5}, and consider the four different
scenarios described below.

{\it Run (a): DI's only, no diffusion}. 
Diffusion is set to zero and  the dynamics of the vector ${\bf p}$ 
(hence of the $\theta$-domains) is dictated solely by the torque.  

{\it Run (b): small diffusivity.}
Diffusion is turned on ($\epsilon=0.01$), but the
torque still dominates the dynamics ($\varepsilon_{HB}=-0.1$).

{\it Run (c): large diffusivity.}
Diffusion ($\epsilon=0.01$) now dominates over
torque ($\varepsilon_{HB}=-0.0001$). 

{\it Run (d): No DI's, diffusion only}.
The system dynamics is entirely dictated by diffusion ($\epsilon=0.01$). 

The parameters for these simulations are collected in Table \ref{table1}.

\begin{table}[htbp] 
  \begin{center}
    \begin{tabular}{|c|c|c|c|c|c|c|}
      \hline
      Run & $\epsilon$ & $V_{HB}$ & $\varepsilon_{HB}$ & $\tau_T /\tau_D$ \\ 
      \hline
      (a) &  $0$       & ON     & $-1\times 10^{-1}$   & $0.0$     \\
      (b) &  $0.01$    & ON     & $-1\times 10^{-1}$   & $10^{-5}$ \\
      (c) &  $0.01$    & ON     & $-1\times 10^{-4}$   & $10^{-2}$ \\
      (d) &  $0.01$    & OFF    & --                   & $\infty$ \\
      \hline
    \end{tabular}
  \end{center}
  \caption{Simulation parameters. 
    The parameter $\epsilon$ fixes the numerical diffusivity 
    (see Eq.~(\ref{diffusivity})); 
    $\tau_T/\tau_D$ is the relative strength of the deterministic torque
    to the diffusive term (see Eq.~\eqref{ratiotau}).
    The drag coefficient is $\gamma=1$, with $C_T=1/\gamma$
    in Eq.~(\ref{totalT}). 
    When the HB potential $V_{HB}$ is switched on (see Eq.~\ref{eq:VHB}), 
    we used the values $\sigma_\theta = \sigma_R = 0.12$, $R_{HB} = \sqrt{2}$
    and we set $\phi(\rho)=1$ 
    to first consider the homogeneous scenario. 
    The value of $\varepsilon_{HB}$ is given in the table.
    The lattice sizes were $N_x=N_y=64$ in all cases.
  } 
  \label{table1}
\end{table}

The equilibrium configurations for $\theta$ in Run (a) and Run (b) are
shown in Fig.~\ref{fig:fig5}, together with the (random) initial
configuration for all runs presented in this section.  As expected,
the directional interactions drive the system towards a disordered
collection of uniform, ordered domains, i.e. within each domain
$\theta$ takes the same value, but domains with different $\theta$
coexist within a disorderd global pattern, in which all four minima
are roughly equally populated (as shown by the histograms).  As
discussed previously, the spatial distribution of the ordered domains
reflects the initial condition, since $\theta$ is attracted to the
minimum closest to its initial value, and once there, it stops
evolving because the system is deterministic (no stochastic
fluctuations).  This corresponds to an arrested-coarsening scenario,
as observed in many slow-relaxing materials, including water.  In the
absence of diffusion, coarsening is virtually quenched, and the
resulting domains are very small and very close to the initial
condition.

\begin{figure}[thbp]
   \includegraphics[width=3in]{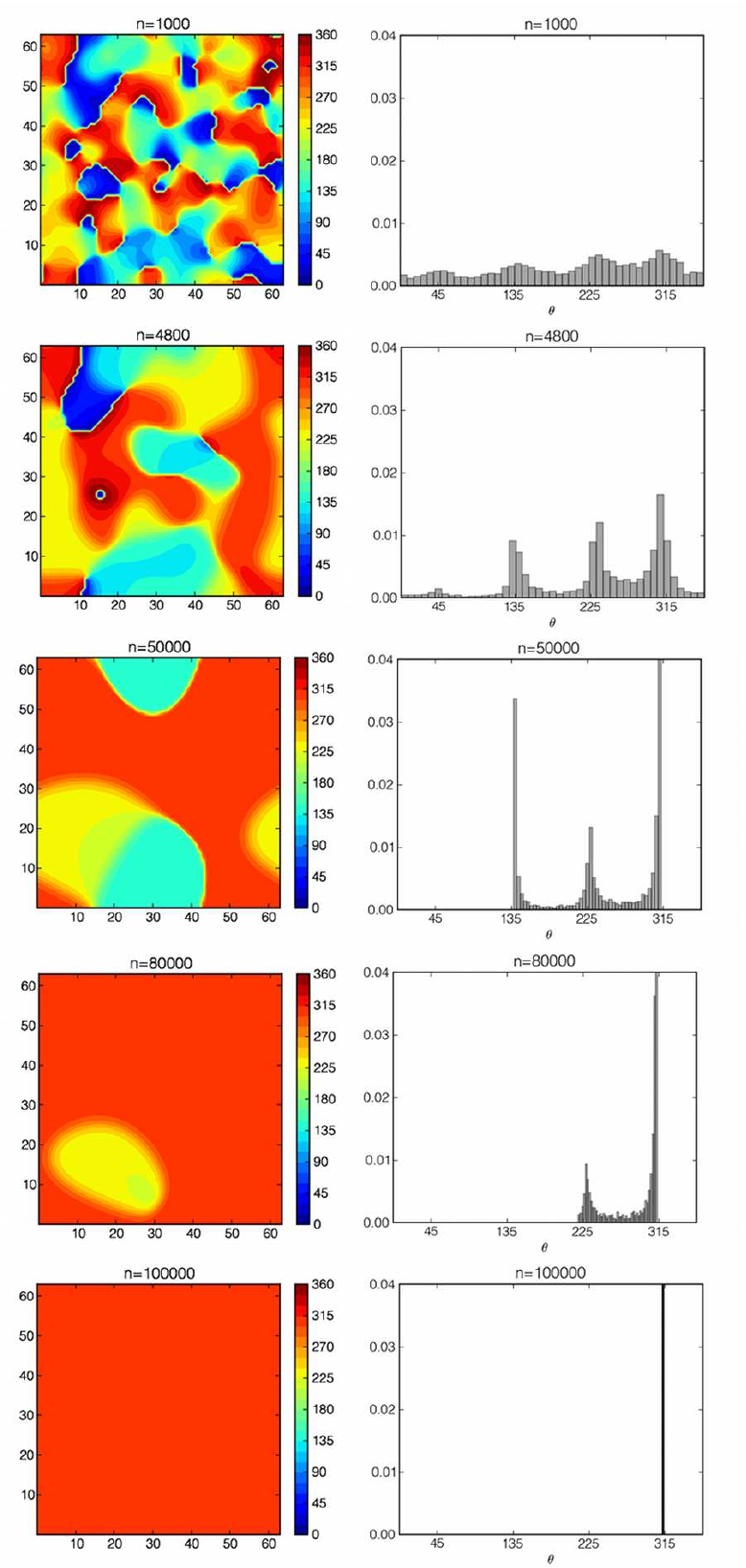}
   \caption{Run (c). Snapshots of $\theta$ (left panels) and
     corresponding histograms (right panels) at different steps $n$ during
     the simulation for a $64^2$ homogenous system with
     $\epsilon_{HB}=-1\times10^{-4}$.  At equilibrium the histogram is
     peaked at the value $\theta=7\pi/4$, which corresponds to one of the
     four minima of $V_{HB}$. The value of the attained minimum depends on
     the initial condition (here, same as in Fig.~\ref{fig:fig5}).  
   }
   \label{fig:fig6}
 \end{figure}

\begin{figure}[htbp]
   \includegraphics[width=3in]{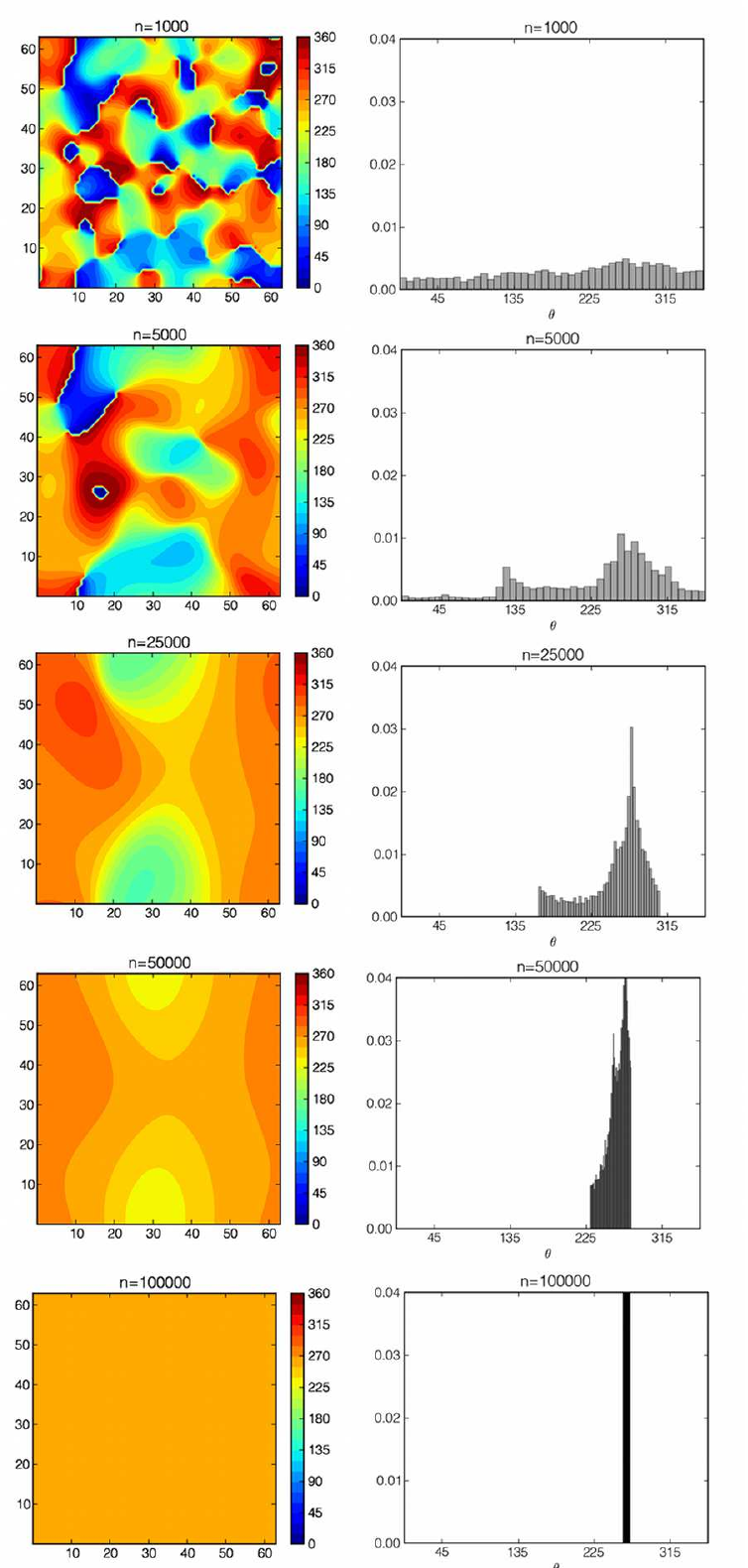}
   \caption{Run (d). Snapshots of $\theta$ (left panels) and corresponding
     histograms (right panels) at different steps $n$ during the
     simulation of a homogenous system of $64^2$ without DI's. 
     The final value of $\theta$ is random, and it depends on
     the initial condition (here, same as in Fig.~\ref{fig:fig5}).
   }
   \label{fig:fig7}
 \end{figure}

When diffusion is turned on, the system experiences additional freedom
to evolve away from the initial condition.  As a result of diffusive
transport, coarsening can now take place, as indicated by the
increased size of the uniform domains.  Since DI's are still dominant,
the global pattern remains disordered. 
As expected, the size of the ordered
domains depends on the relative strength of the HB to the thermal
diffusion.  Whenever diffusion dominates over the HB formation,
coarsening can proceed up to the point where all domains merge into a
single one, i.e.~a uniform value of $\theta$ all over the region
occupied by the fluid.  This behavior is illustrated in
Fig.~\ref{fig:fig6}.  The final value of $\theta$ attained by the
fluid corresponds to one of the four equilibrium angles of $V_{HB}$,
and its specific value is selected by the initial condition.  Also
note that during its relaxation to global equilibrium, the system
first reaches local equilibrium (shown by the peaks in the histogram
around the metastable values of $\theta$), and then the ordered
domains grow and merge, with the largest one absorbing the others.
Although coarsening can proceed up to complete uniformity, the system
still keeps track of HB interactions, in that the final $\theta$
cannot just take any value, but only one of the four degenerate
minima.  This means that, even upon averaging over initial conditions,
the system remains non-ergodic, meaning by this that the probability
distribution function (pdf) $P(\theta)$ is given by a combination of
the four Dirac's deltas, $P(\theta) = \frac{1}{4} \sum_{k=1}^4 \delta
(\theta-\theta_k)$, with $\theta_k = (2k-1) \frac{\pi}{4}$, $k=1,4$.
This scenario lies at other extreme as compared to the no-DI's
situation.  As illustrated in Fig.~\ref{fig:fig7}, diffusion
drives the system to a uniform state, with a relaxation dynamics
similar to the case with weak directional interactions (cfr. with
Fig.~\ref{fig:fig6}).  However, in this case, the intermediate and
final values of $\theta$ can take any value between $[0, 2 \pi]$,
depending on the initial conditions.  On a statistical basis, this is
very different from the case of weak DI's because ensemble averaging
over initial conditions would now produce a uniform pdf, $P(\theta) =
\frac{1}{2 \pi}$.

We wish to emphasize that in the case of diffusion-dominated scenarios, the steady-state
solution of Eq.(23) reads simply as $p_x \equiv p \cos \theta = const$ and
$p_y \equiv p \sin \theta = const$, where $const$ means constant in time {\it and} space.
This means that both the magnitude $p$ and the orientation $\theta$, are uniform in space.
Such uniformity does not correspond to any microscopic ``symmetry-breaking'', but simply reflects
the diffusive structure of the mesoscopic equation of motion of the parameter ${\bf p}$.
Indeed, in the diffusion-dominated scenario, isotropy is recovered by averaging upon initial conditions.

The results discussed in this section show that the present LB model
with directional interactions is capable of sustaining long-lived,
metastable states in the form of a disordered collection of uniform
domains, whose asymptotic size and spatial distribution is controlled
by the relative strength of directional interactions versus diffusion.

\subsection{Heterogeneous scenario: hydrophobic effects}
\label{sec:heterogeneous}
The above results pertain to a homogeneous scenario, with no
hydrodynamic motion and away from solid boundaries. However, most
important water-mediated phenomena take place in the proximity of
solid boundaries, where heterogeneity plays a major role 
\cite{PhysTod,MELCH}.  

In view of future applications, it is therefore important to include
these effects in our model.  To this purpose, we have leveraged the LB
capability to include non-ideal interactions (hydrophobic/hydrophilic)
through density-dependent pseudo-potentials (Shan-Chen model,
Eq.~\eqref{eq:Fww}).
In particular, by adjusting the strength, $G_b$, of attractive
fluid-fluid interactions in the bulk, we can generate an effective
repulsion at walls, since a near-wall fluid layer would experience an
attractive force from the second inner layer in the fluid and no force
(or a smaller one) from the solid layer at the wall, thus resulting in
an effective repulsion from the wall (hydrophobic effect). This
strategy has been used by many authors in LB simulation of confined
microfluids, and shown to give rise to density depletion layers near
the wall (see~\cite{LBmicro} and references therein).  Coupling the
effect of these depletion layers with the density-dependent prefactor
$\phi(\rho)$ of Eq.~\eqref{eq:phiHB}, permits to modulate the fluid
structure as a function of distance from the solid walls.

\begin{figure}[tbhp]
   \includegraphics[width=2.8in]{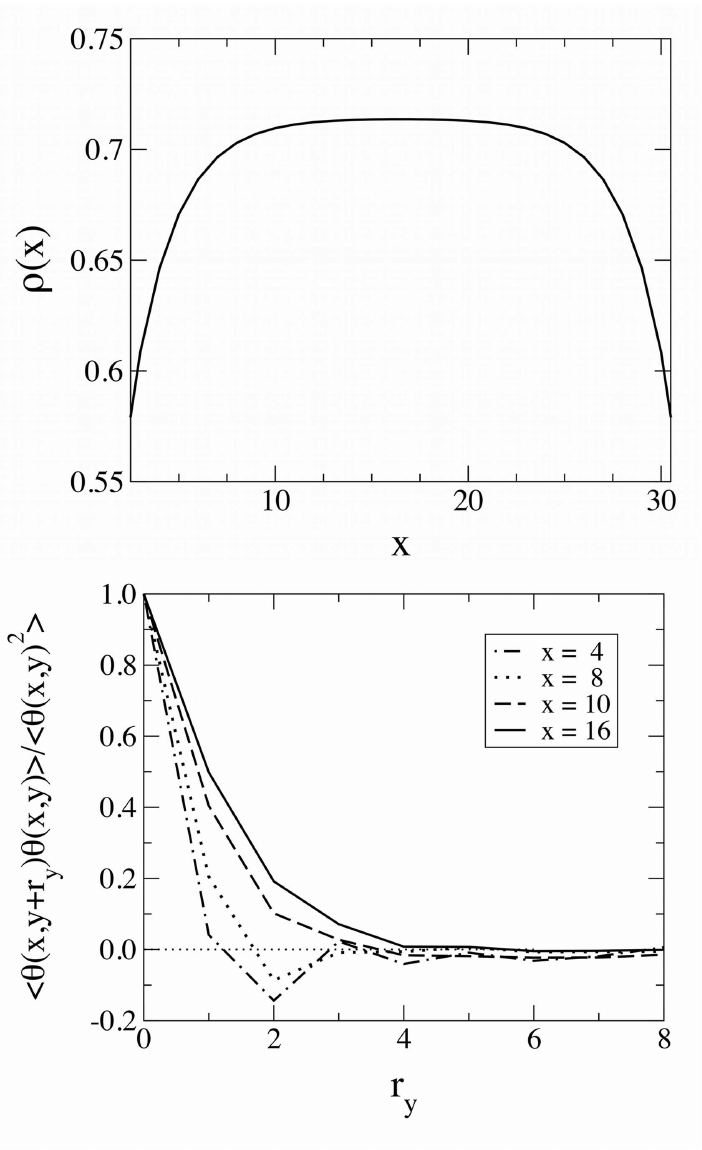}
   \caption{Density profile (top) as a function of the distance from
     the walls located at $x=2.5$ and $x=N_x-1.5$ and correlation function
     of $\theta$ (bottom) as a function of $r_y$ at various locations $x$,
     from close to the wall to the center of the channel.
   }
   \label{fig:fig8}
 \end{figure}

For our tests, we have chosen a channel of size $N_x=32$ and $N_y=128$
with two layers of solid sites perpendicular to the $x$-direction,
located at $x=2.5$ and $x=N_x-1.5$ (for computational reasons, we use
two buffer layers and effective walls lie in between the second buffer
and the first fluid layer).  We used the half-way bounce back scheme
for no-slip boundary conditions, i.e particles exiting the fluid
domain are bounced backed into the fluid with the opposite velocity,
along both normal and tangential directions.  We set the parameter for
the Shan-Chen interaction (see Eq.~(\ref{eq:Fww})) to $G_b=-3.8$, a
value slightly below the critical threshold leading to liquid-vapor
phase separation.  We used the value $\varepsilon_{HB}=-0.1$ in
$V_{HB}$.  The initial condition was set to $\rho({\bf x}) = \ln{2}$
everywhere within the channel.  The density at the wall sites was
fixed to $\rho_{\rm wall}=0.55< \ln{2}$ and we set $\rho_{\rm
min}=\rho_{\rm wall}$.  With these parameters, the maximum density
(reached at the center of the channel, as illustrated in
Fig.~\ref{fig:fig8}), was $\rho_{\rm max}=0.72$, corresponding to a
density depletion ratio of about $25\%$.  
We first consider the static scenario (no net flow) with $\tau_T /\tau_D=10^{-6}$
and $\epsilon=0.001$.

The averaged density profile $\rho(x)=\frac{1}{N_y} \sum_{y=1}^{N_y}
\rho(x,y)$ as a function of the distance from the wall is shown in
Fig.~\ref{fig:fig8} (top).  From this figure, a depletion layer
extending about $\delta = 4 \Delta x$ (4 lattice sites) away from the
wall is well visible.  The width of this layer is of the same order of
magnitude of the interaction range, namely $\sqrt 2 \Delta x$ in the
present nine speed lattice model.

\begin{figure}[tbhp]
   \includegraphics[width=3.25in]{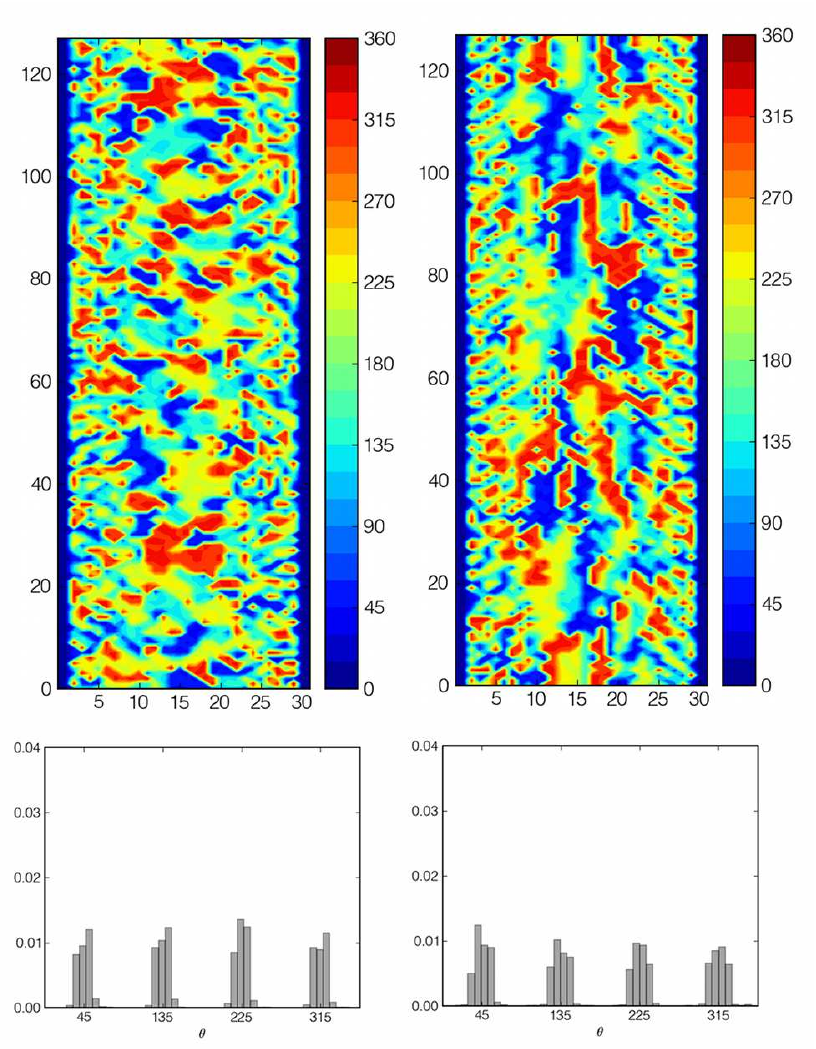} 
   \caption{
     Left panels: Contour plot and corresponding
     histogram of the $\theta$ domains in a heterogeneous system.  The
     histogram is nearly equi-distributed among the four degenerate minima
     of the HB potential.  Right panels: Contour plot and
     corresponding histogram of the $\theta$ domains in a heterogeneous
     system with flow advection. The histogram is still nearly equally
     distributed among the four degenerate minima, but with an enhanced
     spreading due to advective motion.
   }
   \label{fig:fig9}
 \end{figure}

To inspect the ordering effect, we have computed the correlation
function of the $\theta$ domains, namely
\begin{equation}
  C_{\theta}(x; r_y) ={<\theta(x,y+r_y) \theta(x,y)> \over <\theta(x,y)^2>} 
\end{equation}
where the brackets stand for ensemble average over initial conditions and along
the $y$-axis.   
The correlation function 
as a function of $r_y$, at various locations away from the wall,
$x=4,8,10,16$ (in lattice units),  
is shown in Fig.~\ref{fig:fig8} (bottom). 
Near the wall, correlations decay
basically within one lattice site, because the strong effect of the
torque dominates over diffusion effects, quenching the system around
the minima of $V_{HB}$ closest to the initial value of $\theta$.  On
the other hand, at the center of the channel, correlations persist
over larger distances, because diffusive effects lead to ordering
within larger domains.  The contour plot of the $\theta$-domains is
shown in Fig.~\ref{fig:fig9} (left panels), together with the
corresponding histogram.

\subsection{Flowing systems}
\label{sec:flow}
In the Introduction, we have emphasized the crucial role played by
hydrodynamic interactions in water transport phenomena under
confinement \cite{MELCH}.  We have also highlighted the remarkable
capability of LB methods to include hydrodynamic effects in the study
of the dynamics of complex flows, including polar ones \cite{NELIDO,
Cates}, at virtually no extra computational cost.  In this section, we
provide just a qualitative example of such capability, leaving an
in-depth analysis and applications to future publications.  We
consider the heterogeneous scenario discussed in the previous section,
in the presence of a Poiseuille flow between the two confining walls.
Specifically, $u_x=0$ and $u_y(x)=4U_0 (x/W)(1-x/W)$, where $U_0$ is
the centerline speed and $W$ the channel width.  The relevant
dimensionless parameter, measuring the strength of convection versus
diffusion, is the P\'eclet number $Pe=U_0 W/D_p$.  With
$\epsilon=0.001$, hence $D_p=0.001/3$, $W=32$ and $U_0=0.02$, we have
$Pe \sim 2000$, indicating a strong dominance of advective effects
over diffusive ones. However, advection still is three orders of
magnitude slower than DI interactions, which therefore remain the
dominant mechanism. 
Looking at the typical timescales, we have that
$\tau_T /\tau_D=10^{-6}$ and we can estimate 
the ratio $\tau_{adv}/\tau_D$ as $\sim 1/Pe$.
Note that the typical advection time $\tau_{adv}$
takes an intermediate value between the torque and diffusion times.

The main qualitative effect of advective motion, and particularly of
the shear $s_{xy}=\partial u_y/\partial x \sim U_0/W$, is to promote
mechanical deformation of the $\theta$ domains, as one can appreciate
from Fig.~\ref{fig:fig9} (right panels).  This deformation is
accompanied by a mild extra-mixing, as witnessed by the broadening of
the $\theta$ histograms in the bottom panel of the same figure. This
is in line with the qualitative idea of advection as an additional
mixing mechanism, which, in the limit of high P\'eclet numbers,
becomes correspondingly more effective than diffusion in competing
with the ordering effect of directional interactions.
   
As stated above, a quantitative inspection of these complex phenomena
will be left for future studies. Nevertheless,
Fig.~\ref{fig:fig9} conveys an idea of the kind of complexity
that can be tackled by the water-like LB model presented in this work.

\section{Conclusion and Outlook}
\label{sec:outlook}
Summarizing, we have developed a mesoscopic lattice Boltzmann model
for water-like fluids. By water-like, we imply that the fluid includes
directional interactions (DI's) mimicking (some) generic features of
hydrogen-bonds, particularly the tendency to form ordered domains,
with selected bond angles, against the action of thermal noise.  The
competition between the ordering effect of directional interactions
against thermal disorder is shown to lead to the formation of complex
patterns of the orientational order parameter, consisting of a
disordered collection of ordered domains (i.e., in each domain the
order parameter takes a uniform value).  In addition, we have included
non-ideal interactions which permit to realize heterogeneous density
depletion layers near solid walls.  By letting the DI's carry an
inverse density dependence, the model is able to incorporate a
built-in correlation between ordered domains and low density regions,
reflecting the idea of water as a denser liquid in the disordered
state than in the ordered one.

The extension of our model to the three dimensional (3d) case
does not present any conceptual barrier.
The 3d water molecule would be represented by four bonding arms, arranged in
a tetrahedral symmetry~\cite{tetra2}.
This geometry requires two independent orientation angles for each lattice site.
As far as the lattice is concerned, 
a possibility to accommodate the tetrahedral geometry is
the standard cubic 15-speed lattice commonly used in 3d lattice Boltzmann
simulations.

This paper sets a methodological starting basis for LB models of
water-like fluids.  A fully quantitative validation against
microscopic models remains as a task for the future, warranting a
separate investigation on its own.  The model is expected to prove
mostly valuable for the study of nanoscopic water flows with suspended
molecules and/or charged ions, both in standalone \cite{HORBACH}
and/or multiscale scenarios \cite{DELGADO,multiscale}.

\section{Acknowledgements}
Valuable exchange of information with I. Pagonabarraga,
A. Scagliarini, F. Rao and N. Gonz\'alez-Segredo is kindly
acknowledged.  We are grateful to A. Greiner for useful discussions.
MV is supported by the Istituto Italiano di Tecnologia (IIT) under the SEED project grant 
No.259 SIMBEDD.

\bibliographystyle{aip}
\bibliography{lit_water.bib}

\end{document}